# Challenges in assessing Fire Weather changes in a warming climate

Aurora Matteo(1), Ginés Garnés-Morales(2), Alberto Moreno(2), Ribeiro Andreia(3), César Azorín-Molina(4), Joaquín Bedia(5), Francesca Di Giuseppe(6), Robert J. H. Dunn(7), Sixto Herrera(5), Antonello Provenzale(8), Yann Quilcaille(9), Miguel Ángel Torres Vázquez(2), Marco Turco(2)

1. Department of Earth Sciences, University of Pisa, Pisa, Italy
2. Department of Physics, Regional Campus of International Excellence (CEIR) Campus Mare Nostrum, University of Murcia, Murcia, Spain
3. Helmholtz Centre for Environmental Research, UFZ, Leipzig, Germany
4. Centro de Investigaciones sobre Desertificación, Consejo Superior de Investigaciones Científicas (CIDE, CSIC-UV-Generalitat Valenciana), Climate, Atmosphere and Ocean Laboratory (Climatoc-Lab), Moncada, Valencia, Spain
5. Applied Mathematics and Science Computing Department, Universidad de Cantabria, Santander, Spain
6. European Center for Medium-range Weather Forecast (ECMWF), Reading, UK
7. Met Office Hadley Centre, Exeter, UK
8. Institute of Geosciences and Earth Resources, CNR, Pisa, Italy
9. Institute for Atmospheric and Climate Science, Department of Environmental Systems Science, ETH Zurich, Zurich, Switzerland

## Abstract

The Canadian Fire Weather Index (FWI) is widely used to assess wildfire danger and relies on meteorological data at local noon. However, climate models often provide only daily aggregated data, which poses a challenge for accurate FWI calculations in climate change studies. Here, we examine how using daily approximations for FWI95d—the annual count of extreme fire weather days exceeding the 95th percentile of local daily FWI values—compares to the standard noon-based approach for the period 1980–2023.

Our findings reveal that FWI95d calculated with noon-specific data increased globally by approximately 65%, corresponding to 11.66 additional extreme fire weather days over 44 years. In contrast, daily approximations tend to overestimate these trends by 5–10%, with combinations involving minimum relative humidity showing the largest divergences. Globally, up to 15 million km²—particularly in the western United States, southern Africa, and parts of

Asia—exhibit significant overestimations. Among our daily approximation methods, the least biased proxy is the one that uses daily mean data for all variables.

We recommend (i) prioritizing the inclusion of sub-daily meteorological data in future climate model intercomparison projects to enhance FWI accuracy, and (ii) adopting daily mean approximations as the least-biased alternative if noon-specific data are unavailable.

**Plain Language Summary**

Wildfires are becoming a growing concern worldwide due to their increasing intensity and frequency, influenced by climate change. The Fire Weather Index (FWI) is a widely used metric to assess fire danger, but its calculation requires sub-daily meteorological data, such as air temperature and relative humidity at noon. Since climate models often do not provide this information, daily mean data are frequently used as a substitute. In this study, we evaluated the accuracy of these daily approximations compared to a noon-specific calculation by analyzing global FWI95d trends—the annual number of extreme fire weather days—between 1980 and 2023.

Our results show that FWI95d has increased globally by approximately 65% since 1980, equivalent to about 11.66 more extreme fire weather days per year by 2023. Daily approximations consistently overestimate this trend, with increases ranging from 70% to 75%, depending on the variables used. The largest divergences occur in the western United States, southern Africa, and southern Europe, where fire weather trends are most pronounced. Interestingly, the commonly used combination of maximum air temperature and minimum relative humidity yields the greatest overestimation, affecting about 15 million km² of Earth's surface.

To ensure more reliable projections of future wildfire risks, we recommend that climate model intercomparison projects, such as the forthcoming CMIP7, make sub-daily meteorological data available. In the absence of such data, the use of daily mean values for all inputs is the least-biased alternative.

**Key Points**
- FWI95d has increased globally by approximately 65% from 1980 to 2023, translating to 11.66 more extreme fire weather days over 44 years.
- Daily approximations lead to a global overestimation of FWI95d trends, with increases ranging from 70% to 75%. The largest divergences occur in regions such as the western United States and southern Africa.
- To improve fire danger assessments, we recommend (i) making sub-daily meteorological data publicly accessible in future climate model intercomparison projects (e.g., CMIP7) and (ii) relying on daily mean variables as the least-biased alternative if noon-specific data cannot be provided.

## 1. Introduction

The Canadian Fire Weather Index (FWI) is one of the most widely used indicators for evaluating how climatic and meteorological conditions influence wildfire spread once ignition occurs (Taylor and Alexander, 2006; Jones et al., 2022). Relying solely on meteorological inputs—2-meter air temperature, 2-meter relative humidity, 24-hour precipitation, and 10-meter wind speed—the FWI was calibrated for fire weather conditions at the point of maximum air temperature. Because midday local time corresponds to peak fire danger in boreal regions, the FWI formulation uses meteorological data recorded at local noon (Van Wagner, 1987; Vitolo et al., 2020).

However, noon-specific observations are often unavailable in climate model outputs commonly shared through the Earth System Grid Federation (ESGF; https://esgf.llnl.gov/; last accessed 16 January 2025). As a result, researchers have relied on approximations using daily-averaged meteorological data (Flannigan et al., 2013). Still, it remains unclear how much deviations these approximations may introduce.

Answering this question could has significant implications for estimates of changes in landscape flammability, which underpin a substantial body of research conducted in underpinning a substantial body of recent research (e.g., Jolly et al., 2015; Bedia et al., 2015; Jones et al., 2022; Quilcaille et al., 2023). This is particularly relevant as global analyses indicate that high FWI conditions have become increasingly frequent, prolonged, and severe under ongoing greenhouse gas emissions (IPCC AR6 report; IPCC, 2021).

Most studies analysing the projected changes under anthropogenic climate forcings have relied on daily aggregated variables, likely modifying the expected trends. Another concern is that, in the absence of a consensus on the best approach, many studies have adopted different methodological approaches (e.g., substituting air temperature at local noon with maximum or mean daily air temperature), further complicating reproducibility and the comparison of results.

The main aim of this study is to quantify the discrepancy between FWI calculated from daily-averaged versus noon-specific values and to assess the implications this might have. Given that sub-daily data remain sparse in many climate simulations, we address this question through a sensitivity analysis using a reanalysis dataset. Reanalysis datasets integrate observations into atmospheric models to produce physically consistent estimates of climate variables with both spatial and temporal continuity (Kalnay et al., 1996). Such datasets have been widely used in recent fire weather research (e.g., Bowman et al., 2020; Jain et al., 2022;

Jones et al., 2022, 2024). Among these resources, the Copernicus Emergency Management Service FWI dataset (Vitolo et al., 2020)—derived from the ERA5 reanalysis (Hersbach et al., 2020) and developed by the European Centre for Medium-Range Weather Forecasts (ECMWF)—has become a benchmark for evaluating fire danger trends (see e.g., Jones et al., 2022, and references therein).

In this analysis, we calculate the FWI globally from 1980 to 2023 using various daily approximations and compare these results with the FWI dataset from Vitolo et al. (2020), which is based on sub-daily, noon-specific values. Our goal is to determine whether trends derived from approximate inputs differ significantly from those based on noon-specific inputs, as well as to identify approximation strategies that minimize these discrepancies. We assume that our findings can provide a ballpark estimate of the error associated with climate change projections. Therefore, this analysis will inform the reliability and limitations of future global FWI assessments that rely on climate model outputs lacking midday-specific meteorological data.

## 2. Data and Method

We obtain the Fire Weather Index (FWI) dataset (version 4.1; Vitolo et al., 2020) from the Copernicus Emergency Management Service, which provides global data at a spatial resolution of 0.25° × 0.25° with daily temporal resolution, available at https://ewds.climate.copernicus.eu/datasets/cems-fire-historical-v1 (last accessed 16 January 2025). The definition of the FWI requires meteorological data recorded at noon (as in Vitolo et al., 2020); however, climate models typically lack this sub-daily information. To address this limitation, we consider four alternative combinations based on daily variables commonly available from climate models—maximum and mean daily air temperature, mean and minimum daily relative humidity, daily precipitation, and daily mean wind speed—to approximate the FWI. Using these daily variables, we generate four different input combinations (C1 to C4), as summarized in Table 1.

| Comb. | Temp. (°C) | R. Hum. (%) | Precip. (mm) | W. vel. (km h$^{-1}$) |
|---|---|---|---|---|
| C0 | at noon | at noon | 24h* | at noon |
| C1 | DM | DM | 24h | DM |
| C2 | Max | DM | 24h | DM |
| C3 | DM | Min | 24h | DM |
| C4 | Max | Min | 24h | DM |

**Table 1.** Approaches to estimate the FWI. C0 refers to the baseline approach of Vitolo et al. (2020) using the original FWI definition. C1- C4 are the daily-data alternativers replacing noon value. DM= daily mean; Max/Min= daily maximum/minimum. *Precipitation is the 24-hour accumulation ending at noon for C0 and ending at 00 UTC for the other combinations.

Additionally, we perform a sensitivity analysis to assess the impact of individual input variables by systematically replacing each of the four meteorological inputs one at a time with its daily counterpart (e.g., substituting maximum air temperature for noon air temperature while leaving the other variables unchanged). This analysis allows us to isolate the influence of each input on the FWI trends and patterns, providing a clearer understanding of how deviations from the original FWI definition may affect fire weather trends.

To obtain the daily air temperature, relative humidity, precipitation and wind speed values referenced in Table 1, we processed hourly ERA5 data to compute daily minimum, mean, and maximum values. Specifically, for wind speed, we (i) downloaded the hourly 10-meter zonal and meridional wind components (u10 and v10, respectively), and (ii) calculated the daily mean wind speed by averaging the square root of the sum of the squared hourly u10 and v10 components. For relative humidity (RH), we (i) downloaded the hourly 2-meter air temperature, and 2-meter dew point temperature, and (ii) calculated RH using the Magnus formula (Alduchov and Eskridge, 1996).

Then, we computed the FWI from these daily approximations with the *fireDanger* R package (v1.1.0; available at https://github.com/SantanderMetGroup/fireDanger; last accessed 16 January 2025). To validate this method for calculating the FWI, we compared our results - obtained using the same input drivers as Vitolo et al. (2020)- against the original FWI dataset of Vitolo et al. (2020). The results showed no discernible differences (see Figure S1), confirming the reliability of this algorithm for estimating the FWI.

We calculated the FWI starting from 1979 but excluded that year to minimize spin-up effects (Bedia et al., 2018), ensuring that initial conditions did not bias our results. Consequently, our analysis covered the 44-year period from 1980 to 2023. For intercomparison at a resolution typical of global climate models, we bilinearly remapped the data from 0.25º to a 1º x 1º grid, consistent with the IPCC AR6 report (https://github.com/SantanderMetGroup/ATLAS/blob/main/reference-grids/; last accessed 19 December 2024). Comparison tests (Figure S1) confirmed this remapping does not alter our assessment. Following the approach of Quilcaille et al. (2023), we apply a mask based on the ESA Climate Change Initiative land cover dataset from 2016 (ESA-CCI, 2017, 2019). Grid cells with >80% bare areas, water, snow/ice, or sparse vegetation are excluded as areas with infrequent burning (shown in white in subsequent maps). To support the research community, we provide two NetCDF datasets available at https://zenodo.org/records/14964973. The first dataset comprises a mask that delineates regions characterized by infrequent fire occurrences at a 1° resolution. The second dataset consists of a mask designed to identify areas where trend biases are present, based on daily approximation methods. The global spatial mean series was obtained through a spatially weighted average based on the cosine of the latitude, which accounts for the decreasing area of grid cells toward the poles.

We assess trends in extreme fire weather by calculating the FWI95d, defined as the annual number of days when fire weather exceeds the 95th percentile of all daily observations for 1980–2023. This metric is chosen because (i) FWI95d focuses on periods of high fire danger when fire growth is more likely (e.g., Barbero et al., 2014); (ii) many studies have adopted this metric (e.g., Abatzoglou et al., 2019; Jones et al., 2022; Quilcaille et al., 2023); and (iii) as a quantile-based metric, FWI95d minimizes biases in absolute FWI values, enabling a more reliable assessment of trends when using proxies.

We calculated time series slopes via the Theil–Sen estimator (Theil, 1950; Sen, 1968). We then used the modified Mann-Kendall test for serially correlated data, including the variance correction proposed by Hamed and Rao (1998). All tests employed the mmkh function (modifiedmk package; Patakamuri and O'Brien, 2020) in R (R Core Team, 2023). We applied the False Discovery Rate (FDR) method (Benjamini and Hochberg, 1995) for multiple testing correction across the spatial grid.

To determine whether the trend differences between each combination (C1, C2, C3, C4) and the reference dataset (C0) were statistically significant, we followed the same procedure used for the individual time series. Specifically, we (1) computed the difference time series (e.g., C1 – C0), (2) estimated the trend of this difference series using the Theil–Sen estimator, and (3) applied the modified Mann–Kendall test with the Hamed–Rao variance correction to account for serial correlation. We again adjusted for multiple hypothesis testing using the False Discovery Rate procedure. This approach is consistent with Santer et al. (2000), where a difference-based test effectively removes the large-scale variability common to both datasets, making it easier to detect subtle differences between their respective trends. In short, we use the same trend estimation (Theil–Sen) and significance assessment (modified Mann–Kendall) for the difference series as we do for the original series. This ensures consistency and robustness in how we quantify and test the significance of the observed trend differences.

## 3. Results and discussion

The global trend analysis (Figure 1) shows that the baseline FWI95d (C0) increased by 2.65 days per decade between 1980 and 2023. Over 44 years, that amounts to 11.66 additional days, representing a ~65% rise relative to the global average of 18 extreme fire weather days per year. These results align with the findings of Jain et al. (2022) and of Jones et al. (2022), who similarly observed global increases in extreme fire weather. Specifically, Jain et al. (2022) identified decreasing relative humidity and increasing air temperature as key drivers of these trends.

Figure 1 also highlights that none of the daily approximations (C1–C4) preserve the C0 trend exactly, and all combinations overestimate the trend, ranging from 2.86 to 3.06 days per decade. That equates to an approximate increase of 70–75% over the average FWI95d value (Figure 1a). In Figure 1b, the daily approximations exceed the baseline trend by 0.22 to 0.39 days per decade, a difference statistically significant for all combinations.

These results highlight an important caveat for projections of future FWI based on daily-mean meteorological data, as often used in climate models. The reliance on daily averages rather than noon-specific meteorological inputs tends to overestimate the rate of increase in FWI95d globally. It is particularly concerning that combination C4, which uses maximum daily air temperature, minimum daily relative humidity, daily mean wind speed, and daily precipitation, performs the worst among all combinations. C4 is widely accepted as the default approach when subdaily data are unavailable (see e.g. Bedia et al., 2014; Abatzoglou et al., 2019; Quilcaille et al., 2023). However, Figure 1 shows that C4 not only overestimates the trend the most but also introduces the largest deviation from the baseline. These findings underscore the urgency of reassessing the methodologies used to approximate the FWI95d when only daily data are available. For instance, reliance on C4 could result in projections that significantly overstate the risks of future extreme fire weather, emphasizing the need to utilize noon-specific meteorological data wherever possible to improve the accuracy of fire weather projections.

While we later discuss our recommendations in detail, we next explore the spatial differences in trends at the grid scale to better understand why these discrepancies arise and where they are most pronounced.

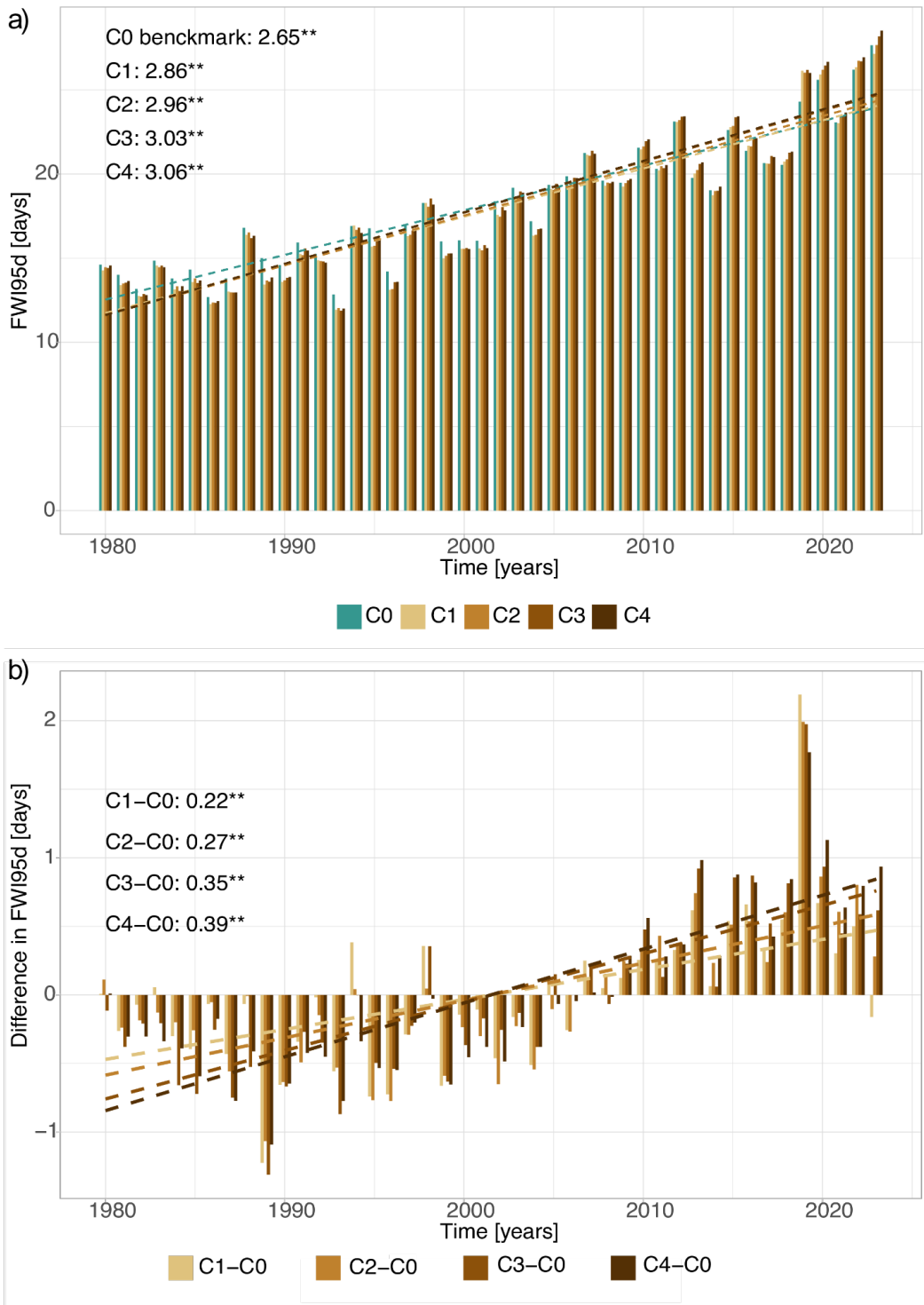

**Figure 1.** Globally averaged FWI95d time series and linear trends from 1980 to 2023. (a) Time series of FWI95d for different input combinations (C0 to C4). (b) Differences in FWI95d trends for daily approximations (C1 to C4) relative to baseline (C0). Trend estimates (in days per decade) are included, with statistical significance indicated (** for p-value < 0.01).

Figure 2 shows the trends in the FWI95d based on our five combinations (C0-C4). The spatial patterns of the trends are notably similar across all combinations and align with previous studies (e.g., Bowman et al., 2020; Jain et al., 2022; Jones et al., 2022), that identified increasing fire weather in several regions globally.

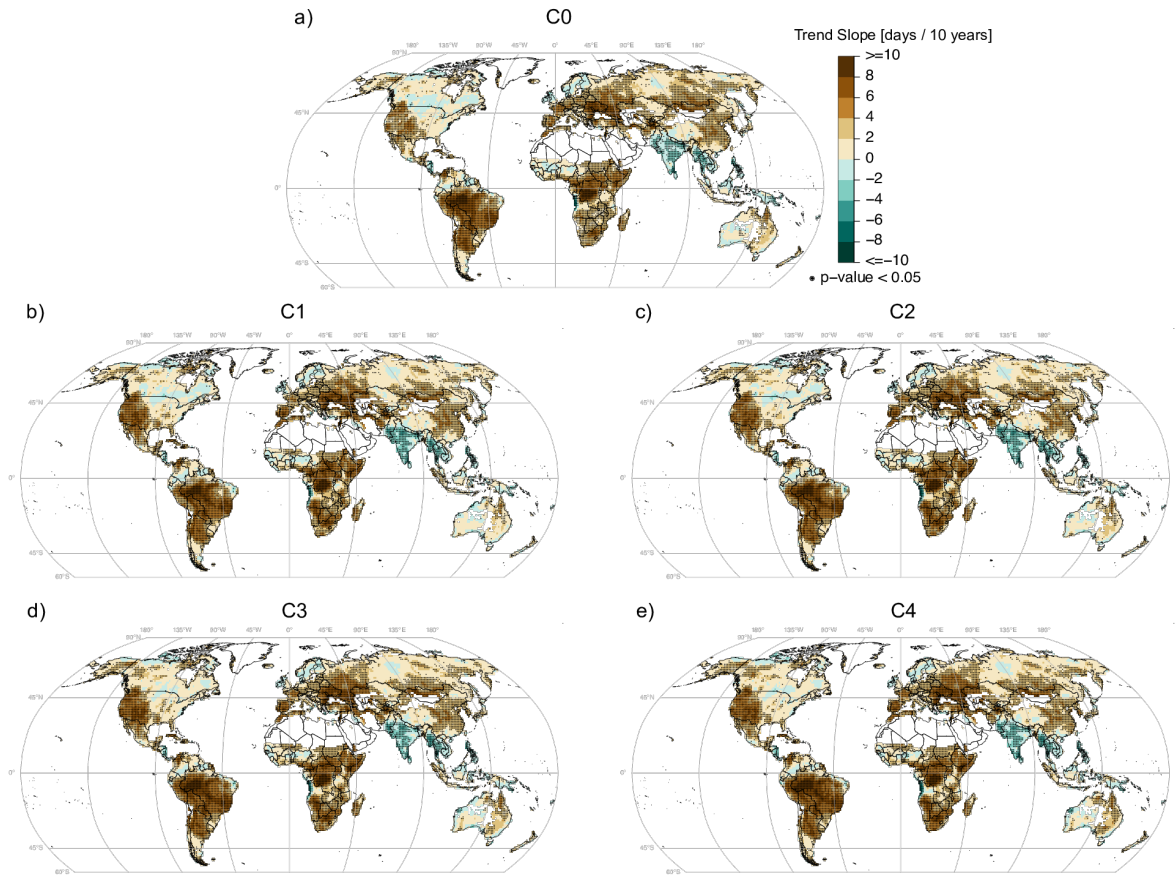

**Figure 2.** Trends in the annual days with FWI values exceeding the 95th percentile (FWI95d) for 1980–2023 for different combinations: a) C0, represents the FWI computed using noon-specific variables, and b) to e) the approximations C1–C4 that use daily mean or min/max values typically available from climate models (Table 1). Regions with significant trends (p-value < 0.05) are outlined with black dots.

Specifically, the western United States shows substantial increases in FWI (Figure 2), underscoring escalating fire weather conditions and wildfire activity in that region -concerns that have been tragically highlighted by recent large-scale fires such as those in the Los Angeles area in January 2025 (Dennison et al., 2014; Abatzoglou and Williams, 2016; Jain et al., 2017; Holden et al., 2018; Turco et al., 2023). Southern and Central Europe, particularly France, Spain, and Portugal, also exhibit pronounced positive trends (Trnka et al., 2021; Hetzer et al., 2024). Similarly, Central and South America, especially Brazil, display significant positive trends (Jones et al., 2022). Africa shows an increase in fire weather, particularly in the central and southern areas. Additionally, significant positive trends are observed in extended areas in Asia, including parts of Turkey and the Middle East. By contrast, India and other parts of South Asia exhibit negative trends, likely tied to increased atmospheric moisture (Jana et al., 2024) and irrigation patterns (Mishra et al., 2020).

Although the trends in Figure 2 appear very similar visually, careful analysis is needed to confirm whether daily combinations conserve or alter the reference trend. Figure 3 shows the trend differences between the four daily approximations (C1–C4) and the baseline (C0) to make it easier to detect divergences in trends. As expected, this highlights regions most sensitive to approximations, revealing potential biases when only daily data are used.

Most grid points do not exhibit statistically significant trends in the differences from the C0 baseline (Figure 3). In general, daily approximations provide reliable FWI95d trends in many regions. Significant positive trends, where the proxy combinations exceed the baseline, are predominantly observed in the western U.S., southern Africa, localized areas in South America, central Africa, the Iberian Peninsula, western Asia, and eastern Australia. In C3 and C4, additional positive trends appear in western Canada. Few areas show negative trends, and these are generally scattered. The extent of these regions varies across combinations (Table 2). It is lower for C1 and C2, which use mean relative humidity, but higher in C3 and C4, which incorporate minimum relative humidity (Table 2).

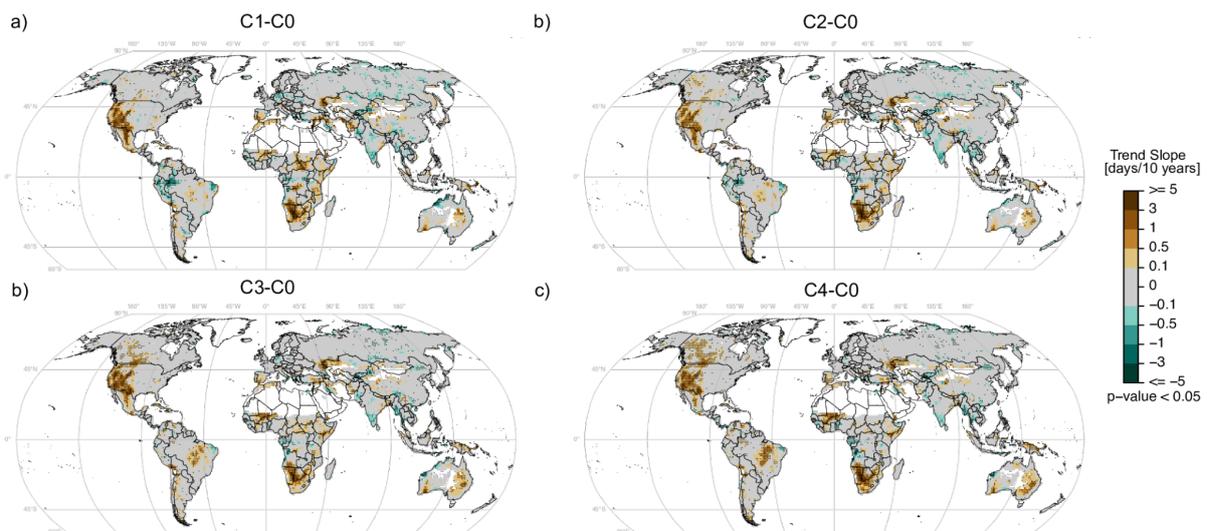

**Figure 3**. Map of the differences in the trends between the FWI95d calculated using four different approximations (C1, C2, C3, C4) and the baseline calculation (C0) for 1980–2023. Brown colours indicate that the FWI95d trend based on approximations overestimates the FWI95d trend based on the baseline values, while green colours indicate underestimation.

Gray colours represent very similar trends (between -0.1 and 0.1 days/10 years). Regions with significant trends (p-value < 0.05) are outlined with black dots.

| Combination | Significant area (M·km²) |
|---|---|
| C1 | 9.76 |
| C2 | 9.44 |
| C3 | 11.65 |
| C4 | 14.82 |

**Table 2**. Areas (million km²) with statistically significant ($p < 0.05$) trend differences from C0.

Based on the results, relative humidity emerges as pivotal in understanding FWI trends when using daily proxies. Figures 4 and S2, which isolate each substituted variable, confirm that substituting relative humidity exerts the strongest effect on FWI trends. In contrast, air temperature, wind speed, or precipitation have comparatively minor impacts (Figure S2). Notably, using mean or minimum RH leads to the largest deviations from C0 in western U.S., southern Africa, and eastern Australia (Figure 4).

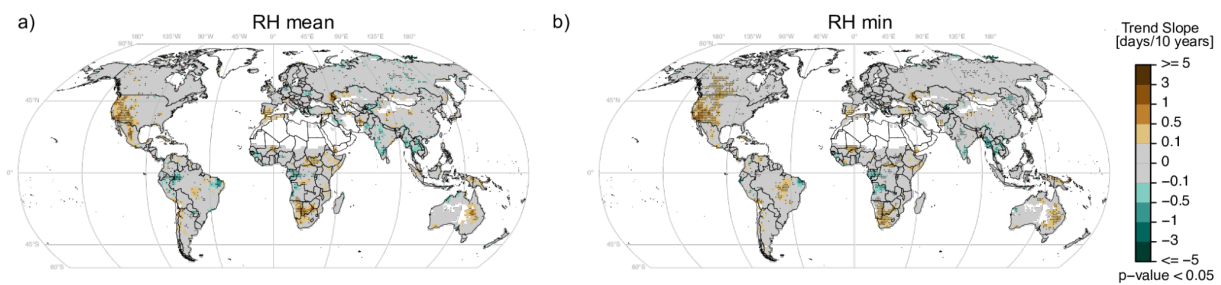

**Figure 4.** Influence of relative humidity proxies. (a) Trends in differences between FWI95d calculated using daily mean RH instead of noon RH, while maintaining the original FWI definition. (b) Similar to (a) but substituting noon RH with daily minimum RH.

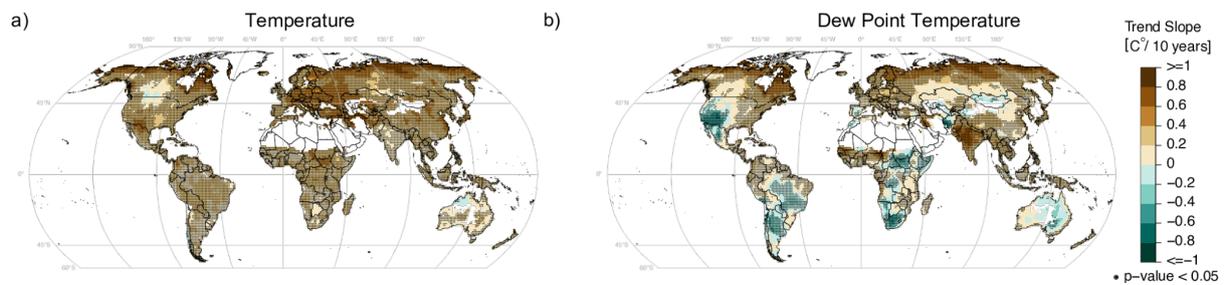

**Figure 5.** Trends in mean air temperature and dew point temperature from 1980 to 2023 in ERA5. (a) Map of the trends in annual mean air temperature. (b) Map of the trends in annual mean dew point temperature. Both panels show trend slopes in °C per decade, with significant trends (p-value < 0.05) marked by black dots.

As highlighted by Jain et al. (2022), the increasing FWI95d globally is primarily driven by rising air temperatures and declining relative humidity. However, although air temperature (T) and RH each influence fire weather, they are not independent: as T increases, RH will decrease unless dew point temperature (Td) also rises. Figure 5 illustrates spatial T and Td trends, showing where RH becomes disproportionately influential in FWI95d. Trends in T are nearly uniform globally, whereas Td trends exhibit considerable spatial heterogeneity. In particular, Td decreases significantly in the western United States, southern Africa, South America, and parts of Australia.

These decreases are likely driven by a combination of processes reflecting the complex interplay between regional climate dynamics, land use, and the water cycle (e.g., Willett et al., 2008; Matsoukas et al., 2011; Vicente-Serrano et al., 2018; Jain et al., 2022). In semi-arid areas, land–atmosphere feedback constrains moisture availability (McKinnon et al., 2021), while deforestation in the Amazon diminishes local evapotranspiration (Barkhordarian et al., 2019). Such decoupling between T and Td results in sharper RH declines, thereby amplifying FWI's sensitivity to humidity changes.

This decoupling of T and Td trends drives sharper declines in RH, which in turn amplifies the sensitivity of FWI to RH changes. Likely, in regions where Td is decreasing, RH becomes more sensitive to even minor fluctuations in moisture availability, exacerbating its influence on FWI trends. Such conditions not only amplify the severity of fire weather but also underscore the potential for RH proxies, such as daily mean or minimum values, to misrepresent the actual fire weather dynamics when noon-specific data are unavailable.

We found that in regions with declining atmospheric moisture, using minimum daily RH (as in C3 and C4) exaggerates the impact of low humidity values on FWI trends. Instead mean daily RH, while less reflective of the noon-specific conditions, inadvertently counteract the tendency for overestimation in FWI trends by providing higher humidity values that can temper the sensitivity of FWI to RH values at noon. Moreover, the observed decreases in Td in these hotspot regions suggest that RH's influence on FWI trends could grow even more pronounced under future climate scenarios, where declining Td may further amplify fire weather risk. This reinforces the importance of accurately representing RH in fire climate models, particularly in regions with declining Td trends, to ensure reliable projections of future fire weather risk.

## 4. Discussion and Conclusions

In our study, we conducted a comprehensive evaluation of the assumptions used in Fire Weather Index (FWI) calculations for climate change applications. Specifically, we assessed the feasibility of approximating noon-specific FWI inputs using daily-mean meteorological data rather than the noon-specific data required by the original FWI definition. We focused on FWI95d, the annual count of days exceeding the local 95th-percentile threshold, for 1980–2023. By comparing the daily approximations against the benchmark by Vitolo et al. (2020), we offered a global perspective of differences and implications for wildfire risk assessment.

Our findings confirm that extreme fire weather is increasing globally: FWI95d has risen by about 65% since 1980, but daily approximations can inflate this figure by an additional 5–10%. This overestimation indicates that future climate projections relying on daily mean data will likely overestimate the rate of global rise in extreme fire weather conditions. This has important implications for the climate modelling community. For future global model experiments, such as the forthcoming Coupled Model Intercomparison Project Phase 7 (CMIP7), we strongly recommend making a greater number of key variables available on a sub-daily basis. This would not only enhance the accuracy of FWI assessments but also improve the reliability of climate change impact studies across various sectors.

Herrera et al. (2013) first demonstrated systematic biases in FWI when using daily means for the Iberian Peninsula, advising the use of noon-specific data for climate projections. Bedia et al. (2014) extended this to Europe and found that combining maximum air temperature and minimum relative humidity (C4) was reliable in representing fire danger. More recently, Quilcaille et al. (2023) provided the first global-scale sensitivity analysis of fire weather indices to input proxies. They compared the performance of C3 and C4 but did not include other combinations tested in this study (C0, C1, C2). Their findings indicated that replacing minimum relative humidity with mean relative humidity consistently reduced FWI during fire seasons, regardless of the time period (1994–2014 or 2081–2100). This suggests that the choice of humidity metric may not drastically alter relative changes in FWI over time. However, here we show that the differences introduced by daily approximations are not constant but accumulate over time, leading to overestimated projections of future fire weather risk. Thus, relatively simple and widely used methods for future projections, like the delta method, cannot fully compensate for these differences because it corrects for the mean biases but does not account for diverging trends.

To summarize, previous studies generally agree that the best proxies for calculating the FWI are noon air temperature paired with daily maximum air temperature, and noon relative humidity paired with daily minimum relative humidity—representing the combination we identify as C4 (Bedia et al., 2014; Abatzoglou et al., 2019; Quilcaille et al., 2023). However, our findings indicate that this combination leads to the highest global overestimation of FWI95d trends, with a 75% increase compared to 65% for the baseline calculation (C0). Moreover, C4 shows the largest area of statistically significant differences, covering approximately 15 million km². Similarly, the combination involving minimum relative humidity (C3) also tends to overestimate FWI trends more than those based on mean relative humidity. Considering these results—and the common limitation of climate models providing daily mean values rather than specific extremes (e.g., minimum relative humidity or maximum air temperature)—we recommend the use of combination C1 (which utilizes daily mean values of air temperature, relative humidity, precipitation, and wind speed) for climate change assessments when noon-specific variables are unavailable.

Our findings also highlight the critical role of relative humidity for fire weather trends, especially in regions experiencing sharply declining atmospheric moisture. In such areas, inaccuracies in RH proxies translate directly into exaggerated FWI estimates. Future research should delve deeper into regional-scale dynamics—especially in data-scarce regions like Africa and South America—to refine local wildfire risk assessments. Overall, we underscore the need for caution when using daily approximations for FWI calculations in climate change studies. By highlighting the factors and regions where discrepancies occur, this work contributes toward more accurate modelling of wildfire risks in a changing climate.

**Acknowledgments**
This work was supported by the project 'Climate and Wildfire Interface Study for Europe (CHASE)' under the 6th Seed Funding Call by the European University for Well-Being (EUniWell). M.T. acknowledges funding by the Spanish Ministry of Science, Innovation and Universities through the Ramón y Cajal Grant Reference RYC2019-027115-I and through the project ONFIRE, Grant PID2021-123193OB-I00, funded by MCIN/AEI/10.13039/501100011033 and by "ERDF A way of making Europe". A.P. acknowledges the support of the EU H2020 project "FirEUrisk", Grant Agreement No.

101003890. Y.Q. acknowledges the support of the EU Horizon Europe project SPARCCLE, Grant Agreement No. 101081369. RJHD was supported by the Met Office Hadley Centre Climate Programme funded by DSIT. C.A-M. received funding from the PROMETEO Ref. CIPROM/2023/38.

**Data Availability Statement**

The datasets used in this study are publicly available. ERA5 hourly variables can be accessed via the Copernicus Climate Data Store at https://cds.climate.copernicus.eu/cdsapp#!/dataset/reanalysis-era5-single-levels. Historical Fire Weather Index data, provided by the Copernicus Emergency Management Service, is available at https://cds.climate.copernicus.eu/cdsapp#!/dataset/cems-fire-historical-v1. The global 1º x 1º land-sea mask grid used in this study is available on GitHub at https://github.com/SantanderMetGroup/ATLAS/raw/main/reference-grids/land_sea_mask_1degree.nc4. Global land cover data consistent with the CCI 1992–2015 map series is accessible at https://www.esa-landcover-cci.org/?q=node/197.

The same meteorological forcings used to calculate the FWI in Vitolo et al. (2020) are available upon request. While this dataset is not publicly accessible, it is not required to reproduce the main results of this study. Additionally, the code used for the analyses in this study is available upon request from the corresponding author.

# Challenges in assessing Fire Weather changes in a warming climate

Aurora Matteo(1), Ginés Garnés-Morales(2), Alberto Moreno(2), Ribeiro Andreia(3), César Azorín-Molina(4), Joaquín Bedia(5), Francesca Di Giuseppe(6), Robert J. H. Dunn(7), Sixto Herrera(5), Antonello Provenzale(8), Yann Quilcaille(9), Miguel Ángel Torres Vázquez(2), Marco Turco(2)

1. Department of Earth Sciences, University of Pisa, Pisa, Italy
2. Department of Physics, Regional Campus of International Excellence (CEIR) Campus Mare Nostrum, University of Murcia, Murcia, Spain
3. Helmholtz Centre for Environmental Research, UFZ, Leipzig, Germany
4. Centro de Investigaciones sobre Desertificación, Consejo Superior de Investigaciones Científicas (CIDE, CSIC-UV-Generalitat Valenciana), Climate, Atmosphere and Ocean Laboratory (Climatoc-Lab), Moncada, Valencia, Spain
5. Applied Mathematics and Science Computing Department, Universidad de Cantabria, Santander, Spain
6. European Center for Medium-range Weather Forecast (ECMWF), Reading, UK
7. Met Office Hadley Centre, Exeter, UK
8. Institute of Geosciences and Earth Resources, CNR, Pisa, Italy
9. Institute for Atmospheric and Climate Science, Department of Environmental Systems Science, ETH Zurich, Zurich, Switzerland

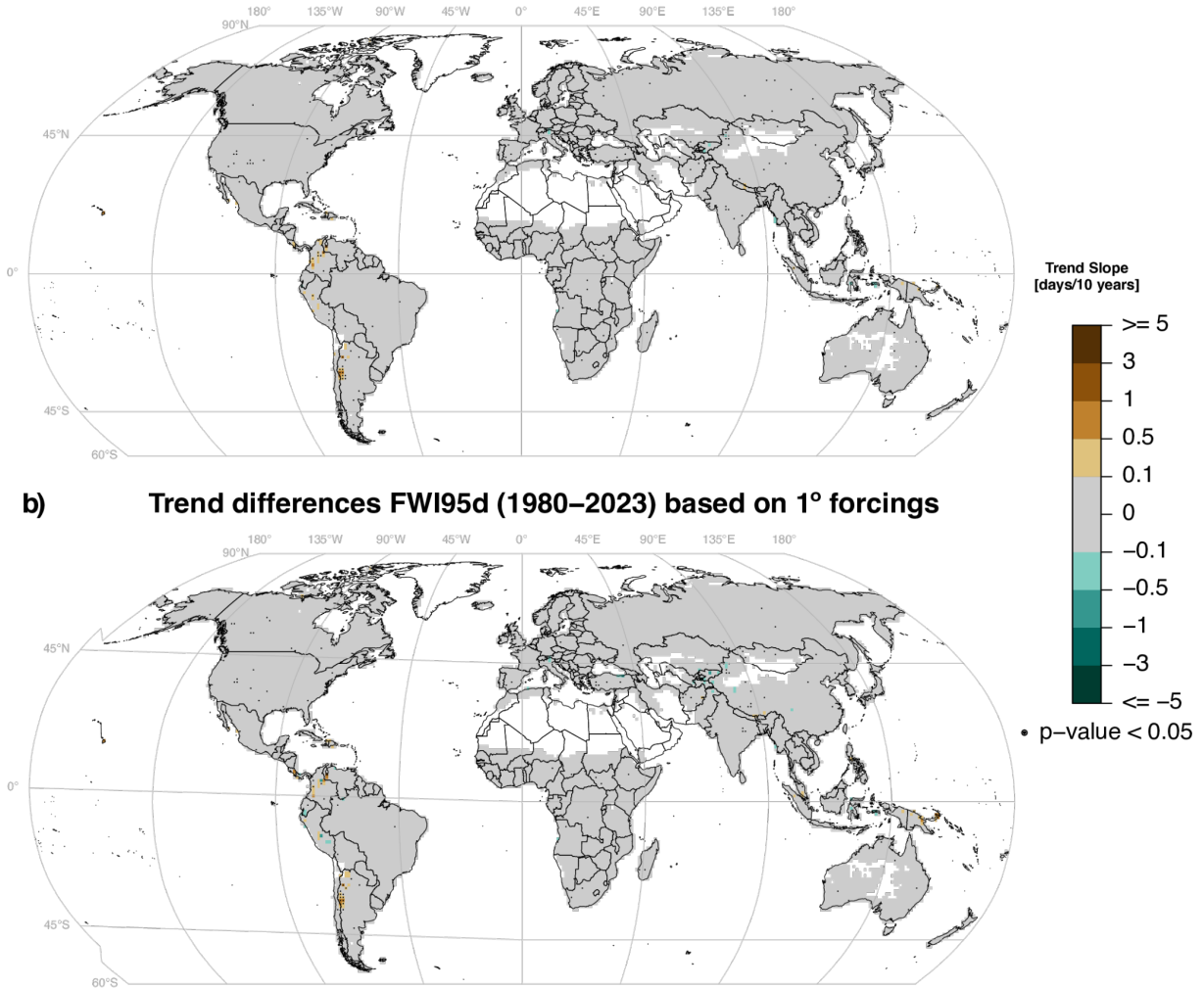

**Figure S1.** Map of the differences in the trends of the FWI95d calculated using a) the R package *fireDanger* (v1.1.0) with the same input drivers as Vitolo et al. (2020) at 0.25° resolution, then interpolated to 1° (baseline calculation C0), and b) using the same input drivers as Vitolo et al. (2020) interpolated to 1° before calculating the FWI. Brown colours indicate that the FWI95d trend based on FWI calculated with *fireDanger* overestimates the FWI95d trend based on FWI from Vitolo et al. (2020), while green colours indicate underestimation. Gray colours represent very similar trends (between -0.1 and 0.1). Areas with significant trend differences (p-value < 0.05) are outlined with black dots.

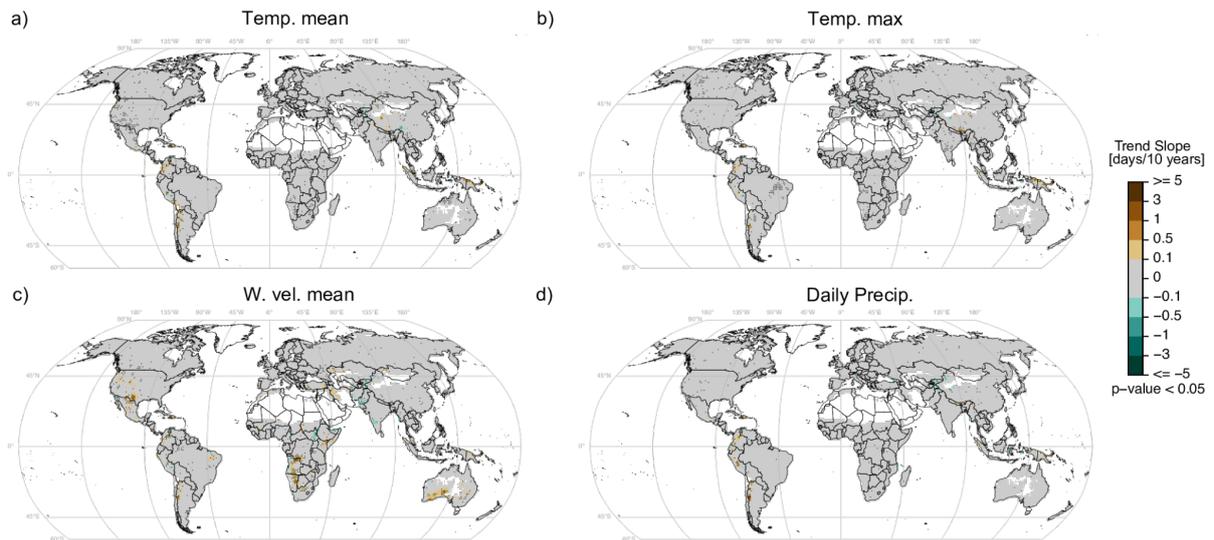

**Figure S2.** Sensitivity analysis of FWI95d trends to relative humidity proxies. (a) Map of the differences in the trends between FWI95d calculated using daily mean air temperature instead of noon air temperature, while maintaining other variables as per the FWI definition. (b) Similar to (a) but substituting noon air temperature with daily maximum air temperature. (c) Similar to (a) but substituting noon wind speed with daily mean wind speed. (d) Similar to (a) but substituting 24-hour accumulated precipitation ending at noon with 24-hour accumulated precipitation ending at 00 UTC. Brown colours indicate that the FWI95d trend based on approximations overestimates the FWI95d trend based on the baseline values, while green colours indicate underestimation. Gray colours represent very similar trends (differences between -0.1 and 0.1 days/10 years). Areas with significant trend differences (p-value < 0.05) are outlined with black dots.